\documentclass[aps,prl,preprint,showpacs,floatfix]{revtex4}
\usepackage{dcolumn}                 
\newcolumntype{w}[1]{D{.}{.}{#1}}
\newcolumntype{.}{D{x}{}{-1}}
\newcommand*{\centt}[1]{\multicolumn{1}{c}{#1}}

\usepackage{times}
\usepackage{nicefrac}
\usepackage{amsmath}
\usepackage{amsfonts}
\usepackage{amssymb}
\usepackage{amsthm}
\newcommand{\bsigma}{\vec{\sigma}}

\newcommand{\bfp}{\vec{p}}

\begin{document}
\preprint{Version 1.0}
\title{Quantum electrodynamics corrections to the $2P$ fine splitting in Li}
\author{Mariusz Puchalski}
\affiliation{Faculty of Chemistry, Adam Mickiewicz University,
             Umultowska 89b, 61-614 Pozna{\'n}, Poland }
\author{Krzysztof Pachucki}
\affiliation{Faculty of Physics, University of Warsaw,
             Ho\.{z}a 69, 00-681 Warsaw, Poland}
\begin{abstract}
We consider quantum electrodynamics (QED) corrections to the 
fine splitting $E(2P_{3/2}) - E(2P_{1/2})$ in the Li atom.  
We derive complete formulas for the $m\,\alpha^6$ and $m\,\alpha^7\,\ln\alpha$ contributions 
and calculate them numerically using highly optimized, explicitly correlated basis
functions. The obtained results resolve disagreement between measurements 
and lay the foundations for investigation of QED effects in light, many-electron atoms.  
\end{abstract}
\maketitle

\section{Introduction}
The inclusion of relativistic effects and correlations between electrons 
in atomic systems gives rise to some fundamental problems related to the many-electron Dirac equation. 
This equation has to include two and more electron-positron pairs to be in accordance with 
Quantum Electrodynamics (QED). Moreover, achieving the correct nonrelativistic limit for 
the energy difference between states of the same orbital momentum is highly problematic \cite{indelicato}. 
For this reason, no accurate relativistic calculation of the lithium $2P_{3/2} - 2P_{1/2}$ splitting 
has been performed so far. For light atomic systems the best approach 
relies on nonrelativistic QED theory, 
where relativistic and QED effects are treated perturbatively, 
while the nonrelativistic Hamiltonian is solved using explicitly
correlated basis sets. This approach has been successfully 
applied to helium \cite{he_yd, he_lev}, lithium \cite{lit_yd, lit_lev}, and beryllium atoms \cite{be_lev}.
The helium fine structure is a very good example; it was calculated to the $m\,\alpha^7$
order and currently serves as one of the most precise QED tests in few-electron systems \cite{he_fs}.     
Conversely, theoretical results for lithium fine structure are much less accurate,
and the various experiments are not always in agreement with each other and with the theory \cite{brown}. 
For example, long standing discrepancies in the isotope shift of the fine structure
have been resolved only recently, and it would appear that the results from previous experiments
and theoretical predictions were both incorrect \cite{brown, sansonetti}.
In this work we aim to significantly improve theoretical prediction of the lithium fine structure
$2P_{3/2} - 2P_{1/2}$ by the complete calculation  of the $m\,\alpha^6$
and $m\,\alpha^7\,\ln\alpha$ contributions.
We derive closed formulas for QED corrections and perform numerical calculations
using explicitly correlated basis sets with Hylleraas and Gaussian functions.  
Such calculations have been performed by Douglas and Kroll
for the helium fine structure of $^3P_J$ levels in Ref. \cite{dk}.
It took 40 years to extend their two-electron $m \alpha^6$ result to an atom
with three electrons, indicating that accurate calculations of QED effects
in many electron systems is a challenging task.
 
\section{Lithium fine structure}
The leading $m\,\alpha^4$ order contribution $E^{(4)}_{\rm fs}$, 
including the all-order electron g-factor, is obtained from the fine-structure Hamiltonian
\cite{bs}
\begin{equation}
E^{(4)}_{\rm fs} = \langle\phi|H_{\rm fs}^{(4)}|\phi\rangle \label{01}
\end{equation}
where
\begin{eqnarray}
H_{\rm fs}^{(4)} &=& \sum_a \frac{Z\,\alpha}{4\,r_a^3}\,\vec \sigma_a\,
\bigl[(g-1)\,\vec r_a\times\vec p_a \bigr]
\label{02} \\ &&+
\sum_{a\neq b}\frac{\alpha}{4\,r_{ab}^3}\,
\vec \sigma_a\bigl[g\,\vec r_{ab}\times\vec p_b
-(g-1)\,\vec r_{ab}\times\vec p_a\bigr]\nonumber
\end{eqnarray}
and $g$ is the electron $g$-factor, and we employ natural units $\hbar = c = m =1$.
The wave function $\phi$ in Eq. (\ref{01}) is a solution of the nonrelativistic 
stationary Schr\"odinger equation corresponding to the $2^2P$ state
\begin{equation}
(H-E)\phi = 0\label{03}
\end{equation}
where
\begin{eqnarray}
H &=& \sum_a \frac{\vec p_a^{\,2}}{2} + V\nonumber \\
V &=& \sum_a -\frac{Z\,\alpha}{r_a} + \sum_{b<a}\frac{\alpha}{r_{ab}}
\end{eqnarray}
The higher-order relativistic correction $E^{(6)}$ is the subject of the
present work. It can be expressed as the sum of the first- and second-order terms
\begin{equation}
E^{(6)} = \langle\phi|H^{(4)}\,\frac{1}{(E-H)'}\,H^{(4)}|\phi\rangle + 
          \langle\phi|H^{(6)}_{\rm fs}|\phi\rangle,\label{05}
\end{equation}
where \cite{bs}
\begin{equation}
H^{(4)} = H^{(4)}_A + H^{(4)}_B + H^{(4)}_C
\end{equation}
\begin{eqnarray} 
H^{(4)}_A &=&\sum_a \biggl\{-\frac{\vec p^{\,4}_a}{8} + \frac{ \pi\,Z}{2}\,\delta^3(r_a) \biggr\}
\\
&& +\sum_{b<a} \biggl\{\pi \, \delta^3(r_{ab}) - \frac{1}{2}\, p_a^i\,
\biggl(\frac{\delta^{ij}}{r_{ab}}+\frac{r^i_{ab}\,r^j_{ab}}{r^3_{ab}}
\biggr)\, p_b^j \biggr\}\,. \nonumber
\end{eqnarray}
\begin{eqnarray} 
H^{(4)}_B &=&\sum_a \frac{Z}{4\,r_a^3}\,
\vec\sigma_a\cdot\vec r_a\times \vec p_a 
\\&&
+\sum_{a \neq b}  
\frac{1}{4\,r_{ab}^3} \vec \sigma_a \,\big( 2\, \vec r_{ab}\times\vec p_b - \vec r_{ab}\times\vec p_a \big)\,.
\nonumber
\end{eqnarray}
\begin{eqnarray}
H^{(4)}_C &=& \sum_{b<a} \frac{\sigma_a^i\,\sigma_b^j}
{4\,r_{ab}^3}\,\biggl(\delta^{ij}-3\,\frac{r_{ab}^i\,r_{ab}^j}{r_{ab}^2}\biggr)\label{08}
\end{eqnarray}
and where $H^{(6)}_{\rm fs}$ is an effective Hamiltonian of order $m\,\alpha^6$.
$H_B^{(4)}$ above coincides with $H_{\rm fs}^{(4)}$ in Eq. (\ref{02}) for $g=2$.
The first derivation of $H^{(6)}_{\rm fs}$ was performed for helium fine structure
by Douglas and Kroll in \cite{dk} using the Salpeter-like approach. 
Numerical evaluation of this splitting has been performed 
to a high degree of precision in \cite{yandrake}
and \cite{yerkrp}. In this work we obtain $H^{(6)}_{\rm fs}$ for lithium 
fine structure using a different approach,
where nonrelativistic expansion is performed in the beginning
at the Lagrangian level, see Ref. \cite{nrqed} and below. 
 
In order to further improve theoretical predictions, the higher-order $m\,\alpha^7$
contribution is not neglected but instead is approximated by the numerically dominating logarithmic part. 
This part is obtained from the analogous result for helium fine structure \cite{fsqed,yerkrp}
by dropping the $\sigma^i\,\sigma^j$ terms because they do not contribute
for states with total electron spin $S=1/2$,
\begin{equation} \label{09}
E^{(7)}_{\rm fs, log} =   \langle H^{(7)}_{\rm fs, log}\rangle 
+2\,\Bigl\langle H^{(4)}_{B} \frac1{(E_0-H_0)'}\,H^{(5)}_{\rm log}\Bigr\rangle
\end{equation}
\begin{equation}
H^{(5)}_{\rm log} = \alpha^2\ln[(Z\,\alpha)^{-2}]\biggl[\frac{4Z}{3}\,\sum_a\delta^3(r_a) 
- \frac{7}{3}\,\sum_{b<a}\delta^3(r_{ab})\biggr]
\end{equation}
\begin{eqnarray}
H^{(7)}_{\rm fs, log} &=&
\alpha^2\ln[(Z\,\alpha)^{-2}] \,\left[
\frac{Z}{3}\,\sum_a
i\,\bfp_a\times\delta^3(r_a)\,\bfp_a\cdot\bsigma_a
  \right. \nonumber \\&& 
-\frac{3}{4}\, \sum_{b\neq a}
i\,\bfp_a\times\delta^3(r_{ab})\,\bfp_a\cdot\bsigma_a 
 \biggr] \,.
\end{eqnarray}
The neglected higher-order corrections are the nonlogarithmic $m\,\alpha^7$ term
and the finite nuclear mass corrections to the $m\,\alpha^6$ contribution.
Corresponding uncertainties are 40 kHz and 15 kHz, 
what together with numerical uncertainties
leads to about 6 ppm accuracy in the Li fine structure.

\section{Spin-orbit Hamiltonian of order $m\,\alpha^6$}
Various approaches are possible to derive $m\,\alpha^6$ correction, and here
we use a variant of nonrelativistic QED, where the effective NRQED 
Lagrangian is obtained by the Foldy-Wouthuysen (FW) transformation of a Dirac
equation. 
\begin{eqnarray}
H_D &=& \vec\alpha\,(\vec p-e\,\vec A) + e\,A^0  \\
H_{FW} &=& e^{i\,S}\,(H_D-i\,\partial_t)\,e^{-i\,S} \nonumber \\ &=&
H_D + i[S,H_D] -\frac{\partial S}{\partial t} +\ldots\nonumber
\end{eqnarray}
We follow here Ref. \cite{nrqed} and additionally introduce
further transformations to simplify the derivation of $m\,\alpha^6$ operators.
The $H_{FW}$ obtained is
\begin{widetext}
\begin{eqnarray}
 H_{FW} &=& e\,A^0 + \frac{1}{2}\,\bigl(\pi^2-e\,\vec\sigma\cdot\vec B \bigr) 
- \frac{1}{8}\,\bigl(\pi^4-e\,\vec\sigma\cdot\vec B\,\pi^2 -
\pi^2\,e\,\vec\sigma\cdot\vec B\bigr)
\nonumber \\ &&
-\frac{1}{8}\Bigl(e\vec\nabla\cdot\vec E_\parallel + e\,\vec\sigma\cdot
\bigl(\vec E_\parallel\times\vec p-\vec p\times\vec E_\parallel \bigr)\Bigr) 
+\frac{e^2}{2}\,\vec\sigma\cdot\vec E_\parallel\times\vec A
\nonumber \\ &&
+\frac{i\,e}{16}\,[\vec\sigma(\vec A\times\vec p-\vec p\times\vec A)\,,\,p^2]
+\frac{e^2}{8}\,\vec E^2_\parallel
+\frac{3}{32}\Bigl\{p^2\,,\,\vec E_\parallel\times\vec p\cdot\vec\sigma\Bigr\}
\nonumber \\ &&
+\frac{5}{128}\,[p^2,[p^2,e\,A^0]]
-\frac{3}{64}\,\Bigl\{p^2\,,\,\nabla^2 (e\,A^0)\Bigr\} 
+\frac{p^6}{16}
\label{14}
\end{eqnarray}
\end{widetext}
where $\vec E_\parallel = -\vec\nabla A^0$.
$H_{FW}$ can be used to derive $H^{(4)}$ as well as $H^{(6)}$. 
Details of such derivation are presented in Ref. \cite{nrqed}.
Here we re-derive $H^{(6)}_{\rm fs}$ with the use of a simpler, but equivalent,
form of $H_{FW}$ in Eq. (\ref{14}). 
Let ${\cal E}_a$ denote 
the static electric field at the position of particle $a$
\begin{equation}
e\,\vec{\cal E}_a \equiv -\nabla_a V =  
-Z\,\alpha\,\frac{\vec r_a}{r_a^3} +\sum_{b\neq a}\alpha\,\frac{\vec r_{ab}}{r_{ab}^3}
\label{17}
\end{equation}
The vector potential at the position of particle  $a$, 
which is produced by all other particles, is
\begin{equation}
e\,{\cal A}^i_{a} \equiv \sum_{b\neq a} \frac{\alpha}{2\,r_{ab}}
\biggl(\delta^{ij}+\frac{r_{ab}^i\,r_{ab}^j}{r_{ab}^2}\biggr)\,
p_b^j + \frac{\alpha}{2}\frac{\bigl(\vec\sigma_b\times\vec
  r_{ab}\bigr)^i}{r_{ab}^3}\,,\label{18}
\end{equation}
Following the derivation in Ref. \cite{nrqed}, 
the higher-order contributions are
\begin{widetext}
\begin{eqnarray}
H_{\rm fs}^{(6)} &=& \sum_a \biggl\{
\frac{3}{16}\, p_a^2\,e\,\vec{\cal E}_a\times\vec p_a \cdot \vec\sigma_a
+\frac{e}{4}\,
\Bigl(2\,p_a^2\,\vec p_a\cdot\vec {\cal A}_{a} 
+ p_a^2\,\vec\sigma_a\cdot\nabla_a\times\vec {\cal A}_{a}\Bigr)
+\frac{e^2}{2}\,\vec\sigma_a \cdot \vec{\cal E}_a\times \vec {\cal A}_{a}
\nonumber \\ &&
+\frac{i\,e}{16}\,\Bigl[\vec {\cal A}_{a}\times\vec p_a\cdot\vec\sigma_a -
\vec \sigma_a\cdot\vec p_a\times \vec {\cal A}_{a}\,, p_a^2\Bigr]
+\frac{e^2}{2}\,\vec {\cal A}_a^{\,2}\biggr\}
+\sum_{b\neq a}\biggl\{
-\frac{i\,\pi\,\alpha}{8}\,\vec\sigma_a\cdot\vec p_a\times\delta^3(r_{ab})\,\vec p_a 
\nonumber \\ &&
+\frac{\alpha}{4}\biggl(
-i\,\biggl[\vec\sigma_a\times\frac{\vec r_{ab}}{r_{ab}},\frac{p_a^2}{2}\biggr]\,
    e\,\vec{\cal E}_b + \biggl[\frac{p_b^2}{2},\biggl[\vec\sigma_a\times\frac{\vec
    r_{ab}}{r_{ab}},\frac{p_a^2}{2}\biggr]\biggr]\,\vec p_b \biggr)\biggr\}
\end{eqnarray}
\end{widetext}
Most of the terms in $H^{(6)}_{\rm fs}$ are obtained in the nonretardation approximation,
which corresponds to replacing the electromagnetic fields in $H_{FW}$
for the particle $a$ with fields that come from all other particles, called $b$. 
The last two are exceptions. The first term under the sum over $a$ and $b$
comes from a Coulomb interaction between electrons, where both electron vertices, 
instead of $e\,A^0$, are of the form
$-\frac{1}{8\,m^2}\bigl[e\vec\nabla\cdot\vec E_\parallel + e\,\vec\sigma\cdot
\bigl(\vec E_\parallel\times\vec p-\vec p\times\vec E_\parallel \bigr)\bigr]$,
and the second term comes from the single transverse photon exchange with the electron vertices
of the form $-e\,\vec p\,\vec A/m$.
\begin{table}[htb]
\caption{$m\,\alpha^6$ and $m\,\alpha^7\,\ln\alpha$ contributions to Li 2P
  fine splitting, in units $m\,\alpha^6$ and $m\,\alpha^7$ correspondingly}
\label{table2}
\begin{tabular}{r@{\hspace{0.5cm}}w{2.8}}
\hline
\hline
\\[-1ex]
  $\langle\phi|H^{(6)}_{\rm fs} |\phi\rangle$                              & -0.202\,1(16) \\
  $\langle\phi|H^{(4)}_B\,\frac{1_{^2S_o}}{E-H}\,H^{(4)}_B|\phi\rangle$      &  0.293\,49 \\
  $\langle\phi|H^{(4)}_B\,\frac{1_{^4S_o}}{E-H}\,H^{(4)}_B|\phi\rangle$      & -0.295\,94(2) \\[2ex]  
  $2\,\langle\phi|H^{(4)}_B\,\frac{1_{^2P}}{(E-H)'}\,H^{(4)}_A|\phi\rangle$ &  0.195\,3(17)\\
  $\langle\phi|H^{(4)}_B\,\frac{1_{^2P}}{(E-H)'}\,H^{(4)}_B|\phi\rangle$    &  0.539\,7(5)\\[2ex]
  $\langle\phi|H^{(4)}_B\,\frac{1_{^4P}}{E-H}\,H^{(4)}_B |\phi\rangle$      & -0.450\,6(2)  \\
  $\langle\phi|H^{(4)}_C\,\frac{1_{^4P}}{E-H}\,H^{(4)}_C |\phi\rangle$      &  0.006\,23\\
  $2\,\langle\phi|H^{(4)}_B\,\frac{1_{^4P}}{E-H}\,H^{(4)}_C|\phi\rangle$    &  0.020\,90\\[2ex]
  $\langle\phi| H^{(4)}_B \,\frac{1_{^2D_o}}{E-H}\,H^{(4)}_B |\phi\rangle$  & -0.751\,13(2) \\ [2ex]    
  $\langle\phi| H^{(4)}_B\,\frac{1_{^4D_o}}{E-H}\,H^{(4)}_B|\phi\rangle$    &  0.733\,27(2)\\
  $\langle\phi| H^{(4)}_C\,\frac{1_{^4D_o}}{E-H}\,H^{(4)}_C|\phi\rangle$    &  0.000\,08 \\
  $2\,\langle\phi|H^{(4)}_B\,\frac{1_{^4D_o}}{E-H}\,H^{(4)}_C|\phi\rangle$  & -0.000\,01 \\[2ex]
  $\langle\phi|H^{(4)}_C\,\frac{1_{^4F}}{E-H}\,H^{(4)}_C|\phi\rangle$       & -0.002\,13 \\[2ex]
  $ E^{(6)}_{\rm fs}$                                                       &  0.087\,1(24)  \\[2ex]
  $\langle\phi|H^{(7)}_{\rm fs, log} |\phi\rangle$                          &  -0.736\,38   \\
  $2\,\langle\phi|H^{(4)}_B\,\frac{1_{^2P}}{(E-H)'}\,H^{(5)}_{\rm log}|\phi\rangle$ &   1.783\,9(4) \\ [2ex]
  $ E^{(7)}_{\rm fs, log}$                                                        &   1.047\,5(4)  \\[1ex]
\hline
\hline
\end{tabular}
\end{table}
The second-order contribution in Eq. (\ref{05}) is split into parts coming from intermediate
states of the specified angular momentum and the spin. These parts 
are defined in Table \ref{table2}. 
Most of them can be calculated as they stand. 
Only the matrix elements involving $H_A^{(4)}$ and $\delta^3(r_a)$
need special treatment due to the high singularity of these operators.

\section{Spin reduction of matrix elements}
The wave function of the $^2$P state in the three-electron system is represented as
\begin{equation}
\Phi^i = \frac{1}{\sqrt{6}}\,{\cal A}\big[\phi^i(\vec r_1,\vec r_2,\vec r_3)\,
[\alpha(1)\,\beta(2)-\beta(1)\,\alpha(2)]\,\alpha(3)\big]\,,\label{29}
\end{equation}
where $\cal A$ denotes antisymmetrization and 
$\phi^i(\vec r_1,\vec r_2,\vec r_3)$ is a spatial function with Cartesian
index $i$ that comes from any of the electrons coordinate, $\phi^{i} = r_a^i \; \phi$.
The normalization we assume is
\begin{eqnarray}
1 &=& \sum_i \langle\Phi'^i|\Phi^i \rangle \nonumber \\ &=& \sum_i
\bigl\langle \phi'^{\,i}(r_1,\,r_2,\,r_3)|{\cal P}
             [c_{123}\,\phi^i(r_1,r_2,r_3)] \bigr\rangle 
\end{eqnarray}
where $\cal P$ denotes a sum of all permutations of 1,2,3 subscripts.
The $^2$P$_{1/2}$ and $^2$P$_{3/2}$ wave functions are constructed using
Clebsch-Gordon coefficients. Expectation values with these wave functions
can be reduced to spatial expectation values with algebraic prefactor
for $J=1/2,3/2$, accordingly; i.e. the first-order matrix elements take the form 
\begin{widetext}
\begin{equation}
\langle\Phi'|O|\Phi \rangle = 
\{1,1\} \, \bigl\langle \phi'^{\,i}(r_1,\,r_2,\,r_3)|Q\, {\cal P} [c_{123}\,\phi^i(r_1,r_2,r_3)] \bigr\rangle
\end{equation}
\begin{equation}
 \langle \Phi'|\sum_a \vec \sigma_a \cdot \vec Q_a |\Phi \rangle =   
\{1,-1/2\} \, i\,\epsilon^{ijk} \bigl\langle\phi'^{\,i}(r_1,\,r_2,\,r_3)|
\sum_a Q_a^j\,{\cal P} \Big[c^{Fa}_{123}\,\phi^k(r_1,r_2,r_3)\Big] \Big \rangle 
\end{equation}
\begin{eqnarray}
 \langle \Phi'|\sum_{a \neq b} \vec\sigma_a\times \vec\sigma_b\cdot \vec Q_{ab}|\Phi \rangle &=&
 \{1,-1/2\} \, (-2\,\epsilon^{ijk}) \bigl\langle \phi'^{\,i}(r_1,\,r_2,\,r_3)|
\nonumber \\&&
 \sum_{ab=12,23,31} \,(Q_{ab}^{j}-Q_{ba}^{j})\,{\cal P} \Big[c^{F_1}_{123}\,\phi^k(r_1,r_2,r_3)\Big] \Big \rangle 
\end{eqnarray}
\end{widetext}
The second-order matrix elements can also been reduced
to the spatial ones, with different prefactors similarly to those above. 
\begin{table}[!hbt]
\renewcommand{\arraystretch}{1.0}
\caption{Symmetrization coefficients in spatial matrix elements}
\label{table6}
\begin{ruledtabular}
\begin{tabular}{lrrrr}
$(k,l,m)$  & $c_{klm}$ & $c^{F1}_{klm}$ & $c^{F2}_{klm}$ & $c^{F3}_{klm}$ \\
\hline 
 \\
 $(1,2,3)$    &   2   &    0    &    0    &     2   \\
 $(1,3,2)$    &  -1   &    1    &   -1    &    -1   \\
 $(2,1,3)$    &   2   &    0    &    0    &     2   \\
 $(2,3,1)$    &  -1   &   -1    &    1    &    -1   \\
 $(3,1,2)$    &  -1   &    1    &   -1    &    -1   \\
 $(3,2,1)$    &  -1   &   -1    &    1    &    -1   \\
  \end{tabular}
\end{ruledtabular}
\end{table}

\section{Numerical calculations}
The spatial function is represented as a linear combination of
the Hylleraas \cite{wang} 
\begin{equation}
\phi =  e^{-\alpha_1 r_1 -\alpha_2 r_2 -\alpha_3 r_3}\, r_{23}^{n_1}\, r_{31}^{n_2}\, r_{23}^{n_3}\,
                          r_1^{n_4}\,r_2^{n_5}\,r_3^{n_6}
\end{equation}
or of the Gaussian functions \cite{gauss} 
\begin{equation}
\phi = e^{-\alpha_1 r_1^2 -\alpha_2 r_2^2 -\alpha_3 r_3^2 -\alpha_{12} r_{12}^2 -\alpha_{13} r_{13}^2 -\alpha_{23} r_{23}^2}
\end{equation}
In the Hylleraas basis we use 6 sectors with different values of nonlinear parameters $w_i$
and a maximum value of $\Omega = n_1+n_2+n_3+n_4+n_5 = 13$, details are in \cite{lit_fine_yd, lit_fine}.
In Gaussian basis we use $N=256, 512$, $1024$, and $2048$ functions 
with well-optimized nonlinear parameters for each basis function, separately.
The accuracy achieved for nonrelativistic energies is about $10^{-13}$ 
in Hylleraas and $10^{-11}$ in Gaussian bases.
The first-order matrix elements involving the Dirac $\delta$-function are
calculated with Hylleraas basis; all other operators are calculated using
Gaussians. Numerical results for the extrapolated value of $\langle H^{(6)}_{\rm fs}\rangle$
are presented in Table \ref{table2}, and the achieved precision is about $10^{-3}$.
The evaluation of second-order matrix elements is much more demanding.
They are obtained using the Gaussian basis, as follows.
The resolvent $1/(E-H)$ for each angular momentum is represented in terms of 
functions with the appropriate Cartesian prefactor.
Nonlinear parameters for intermediate states are optimized for each symmetric
matrix element. For the asymmetric matrix elements,
the basis is combined from two corresponding symmetric ones.
The most computationally demanding matrix elements were these, 
which involve $H_A^{(4)}$ and $H_{\rm log}^{(5)}$ operators, 
and they are transformed to the regular form by the following transformations
\begin{equation}
H^{(4)}_A = [H^{(4)}_A]_r + 
\bigg\{\sum_a\frac{Z}{4\,r_a} - \sum_{b<a} \frac{1}{2\,r_{ab}},E-H \bigg\}\,, \label{16} 
\end{equation}
\begin{equation}
4 \pi\, \delta^3(r_a) = 4 \pi \, [\delta^3(r_a)]_r - \bigg\{\frac{2}{r_a} ,E-H \bigg\}\,.
\end{equation}
The resulting second order matrix elements became less singular
and can readily be evaluated numerically.
Numerical results for matrix elements are summarized in Table \ref{table2}.
The achieved precision is of order $10^{-3}$ and better; similarly to the first-order matrix elements.
Moreover, we observed significant cancellations between $S=1/2$ and $S=3/2$ intermediate
states, and between the first- and second-order terms.
The final numerical result for the $m\,\alpha^6$ contribution
in Table \ref{table2} is quite small but larger than the hydrogenic value $5/256 =
0.019531$, as it should be. 
Regarding the $m\,\alpha^7$ contribution, the second-order term is numerically
dominant, and the contribution from $H^{(7)}_{\rm fs, log}$ is more than twice smaller.
Altogether, this correction is only 10 times smaller than the $m\,\alpha^6$ contribution
and is significant in comparison to the accuracy of experimental values. 

\section{Summary}
We have performed accurate calculations of the fine structure in Li
using the nonrelativistic QED approach. Relativistic and QED corrections 
are represented in terms of effective operators and are calculated
using a highly accurate nonrelativistic wave function. 
Numerical results are summarized in Table \ref{table5}. 
The obtained theoretical results for the $^{6,7}$Li fine structure are in an agreement
with the recent experimental values of Ref. \cite{brown} and also with
Refs. \cite{brog, orth}, but are in disagreement with all the other ones.
This demonstrates the capability of NRQED theory and the numerical approach
based on explicitly correlated functions in achieving high-precision
predictions for energies and energy splittings in light, few-electron atoms.
\begin{table*}[!hbt]
\renewcommand{\arraystretch}{1.0}
\caption{Fine splitting of 2P-states in Li isotopes in MHz. 
         $\delta E_{\rm fs}$ is the hyperfine mixing correction. The
         uncertainty of $E_{\rm fs}$(theo) comes mainly from numerical
         inaccuracies of $E_{\rm fs}^{(6,0)}$ and from the neglect
         $E_{\rm fs, nolog}^{(7,0)}$, which is estimated by $25\%$ of $E_{\rm fs, log}^{(7,0)}$}
\label{table5}
\begin{ruledtabular}
\begin{tabular}{rw{6.7}w{6.7}l}
         & \centt{$^6$Li} & \centt{$^7$Li} &\centt{Ref.}\\
$E_{\rm fs}^{(4,0)}$             & 10\,053.707\,2(83)  & 10\,053.707\,2(83) &\cite{lit_fine_yd,lit_fine} \\
$E_{\rm fs}^{(4,1)}$             &      -2.786\,8(6)   &      -2.389\,1(5)  &\cite{lit_fine_yd,lit_fine} \\
$E_{\rm fs}^{(6,0)}$             &       1.63(5)       &       1.63(5)      &  \\
$E_{\rm fs, log}^{(7,0)}$         &       0.15         &       0.15      &  \\
$\delta E_{\rm fs}$             &       0.012\,17      &       0.159\,16    &\cite{lit_fine}\\
$E_{\rm fs}$(theo)              & 10\,052.72(6)        & 10\,053.25(6) \\[1ex]
$E_{\rm fs}$(exp)               & 10\,052.779(17)      & 10\,053.310(17) & Brown \cite{brown}\\
                              & 10\,052.76(22)       & 10\,053.24(22) & Brog \cite{brog}\\
                              &                      & 10\,053.184(58) & Orth \cite{orth}\\
                              & 10\,052.964(50)      & 10\,053.119(58) & Noble \cite{noble}\\
                              & 10\,052.044(91)      & 10\,052.37(11) & Walls \cite{walls}\\
                              & 10\,052.862(41)      & 10\,051.999(41) & Das \cite{das}\\
\end{tabular}
\end{ruledtabular}
\end{table*}

\section*{Acknowledgments}
The authors acknowledge the support of NCN grants 2012/04/A/ST2/00105,
2011/01/B/ST4/00733, and by PL-Grid Infrastructure.

\end{document}